\documentclass[aps,prb,onecolumn,showpacs,superscriptaddress,groupedaddress]{revtex4-1}
\usepackage[dvips]{graphics,graphicx}
\usepackage{epstopdf}
\usepackage{bbold}
\usepackage[usenames, dvipsnames]{color}

\begin{document}
\title{Propagating bound states in the continuum in dielectric gratings}

\author{E.N. Bulgakov$^{1,2}$}
\author{D.N. Maksimov$^{1,2}$}
\author{P.N. Semina$^{1}$}
\author{S.A. Skorobogatov$^{1}$}

\affiliation{$^1$Reshetnev Siberian State University of Science and Technology, 660037, Krasnoyarsk,
Russia\\
$^2$Kirensky Institute of Physics, Federal Research Center KSC SB
RAS, 660036, Krasnoyarsk, Russia}
\date{\today}



\begin{abstract}
We consider propagating bound states in the continuum in dielectric gratings.
The gratings consist of a slab with ridges periodically arranged
ether on top or on the both sides of the slab. Based on the Fourier modal approach we recover the
leaky zones above the line of light to identify the
geometries of the gratings supporting Bloch bound states propagating in the direction perpendicular to
the ridges. Most importantly, it is demonstrated that if a two-side grating possesses either mirror or
glide symmetry
the Bloch bound states are stable to variation of parameters as far as the above symmetries are preserved.
\end{abstract}

\maketitle

\section{Introduction}

High contrast dielectric gratings (DG) have become an important instrument in optics
with various application including high-Q resonators and focusing reflectors \cite{Zhou08,Fattal10,Lu10,Karagodsky11,Chang-Hasnain12,Ko16}.
In this paper we address the capacity of DGs to host optical bound states in the continuum (BICs),
i.e. localized eigenmodes of Maxwell's equations with infinite Q-factor
embedded into the continuous spectrum of the scattering
states \cite{Hsu16}. In the recent past the optical BICs were experimentally observed in all-dielectric
set-ups with periodically varying permittivity \cite{Plotnik,Weimann13,Hsu13,Foley14a,Vicencio15,Sadrieva17,Xiao17}.
Nowadays, the optical BICs are employed to engineer high-Q resonators for enhancement of light-
matter interactions with applications to narrow-band transmission filtering \cite{Foley14},
lasing \cite{Kodigala17}, and second harmonics generation \cite{Wang18}.

 Depending on the spacial extension of the light holding structure one can identify
three classes of BICs. If the structure is confined in all three dimensions, a perfectly localized optical mode
with infinite Q-factor can be found in spherical dielectric particles coated with zero-epsilon metamaterial
\cite{Monticone14,Li17a}. One the other hand, if the structure is infinitely extended in
one spacial dimension the above condition on the dielectric permittivity is lifted allowing for BICs in
periodic arrays of lossless high-index dielectric elements such as spheres \cite{Bulgakov2015,Bulgakov17}
and discs \cite{Bulgakov17b}. Notice that in the latter case light is localized only in
two dimensions.

The third class of BIC supporting systems are planar structures infinitely extended in two dimensions.
 They include
perforated slabs  \cite{Shipman2005,Hsu13,Zhen2014,Mocella15,Ni16,Li16,Penzo17},
arrays of rods \cite{Venakides03,Marinica08,Bulgakov2014,Yuan17a,Yuan17b,Hu18}, arrays of
rectangular bars \cite{Zhen2014,Blanchard16,Cui16a,Foley14a,Wang16b,Ni16,Taghizadeh17,Sadrieva17},
and gratings \cite{Marinica08,Yoon15,Cui16a,Monticone17,Wang16}.
Here, we consider Bloch BIC in DGs, i.e.
localized modes propagating above the line of light in the plane of the structure
\cite{Venakides03,Ndangali2010,Yang2014,Ni16,Gao16,Hu17}.
Such BICs can be contrasted to symmetry protected standing waved BICs
\cite{Pacradouni00,Foley14a,Mocella15,Wang16,Cui16a}
that are symmetrically mismatched with the
outgoing wave allowed in the ambient medium.
The Bloch BICs considered here not only provide access to light localization
and concurrent effects of resonant enhancement and frequency filtering, but also allow for light guiding above the line
of light paving a way for multifunction optical elements which steer the flow of light
harvested from the ambient medium \cite{Bulgakov16}.

\section{System overview}
\begin{figure}[t]
\begin{center}
\includegraphics[width=.49\textwidth, height=.245\textwidth,trim={5.6cm 11.5cm 1.0cm 11.0cm},clip]{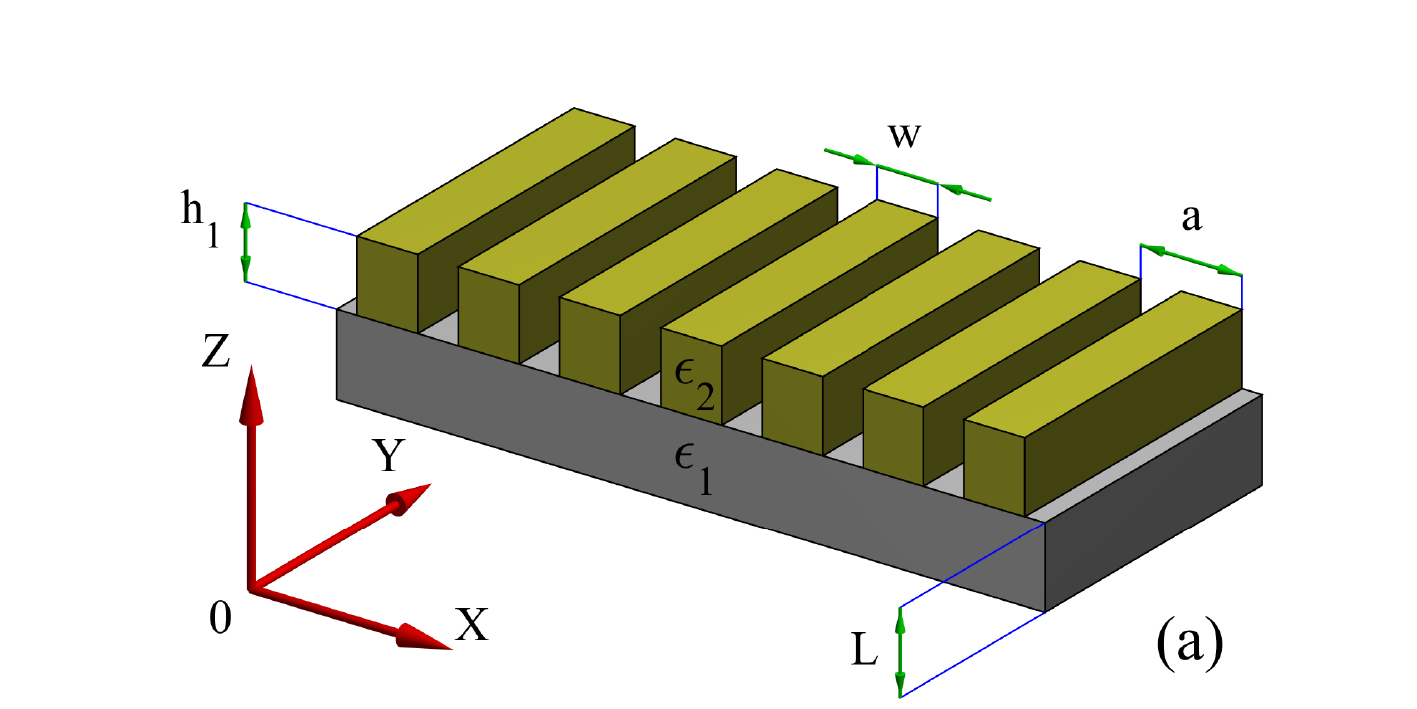}
\includegraphics[width=.49\textwidth, height=.245\textwidth,trim={4.5cm 11.0cm 1.0cm 10.5cm},clip]{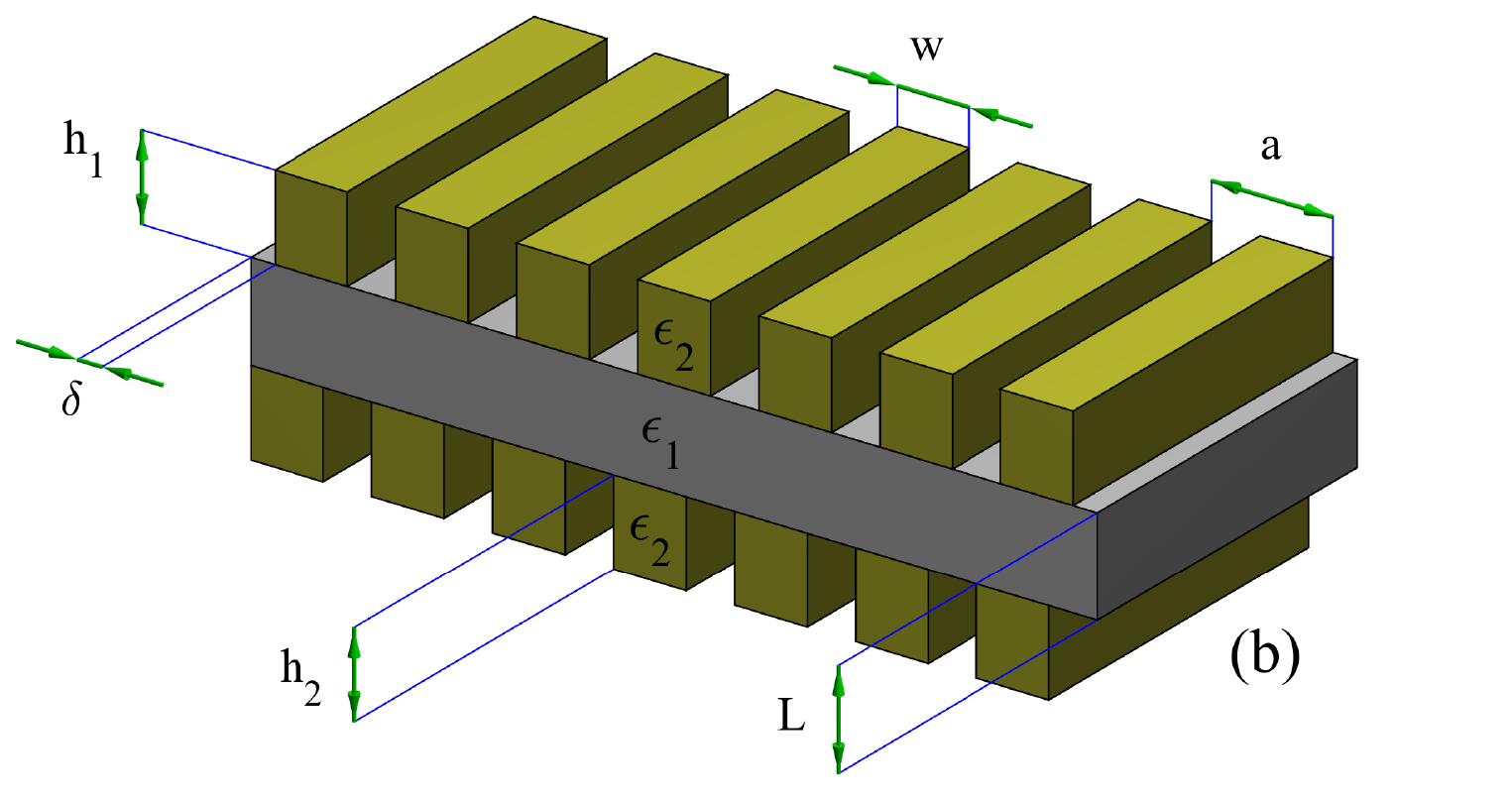}
\caption{Dielectric gratings: (a) One-sided grating; (b) Two-sided grating.}\label{Fig1}
\end{center}
\end{figure}

The simplest DG supporting BICs is sketched in Fig. \ref{Fig1}(a). It consists of a slab substrate of
thickness $L$ made of a dielectric material with permittivity $\epsilon_1$. Dielectric ridges of width $w$
with permittivity $\epsilon_2$ are placed on the top of the slab with period $a$ in the $x$-direction. The ridges
are parallel and infinitely extended
along the $y$-axis. The whole structure is immersed into the ambient medium with $\epsilon_0=1$.
In what follows the thickness of the topside ridges are designated by $h_1$.
In a more generic case of the two-sided DG shown in Fig. \ref{Fig1}(b), the
ridges are also placed on the underside of
the slab. The underside ridges are positioned with the same period $a$ to preserve
the periodicity of the structure as a whole.
The topside and underside ridges are shifted with respect to each other by distance $\delta$.
The thickness of the underside ridges is designated by $h_2$.

Due to the system's translational symmetries the spectral parameters of the eigenmodes are linked
through the following dispersion relationship \cite{Popov12}
\begin{equation}\label{spectrum}
k_0^2=k_{x,n}^2+k_z^2+k_y^2, \ k_{x,n}=\beta-2\pi{n}/a,
\end{equation}
where $k_0$ is the vacuum wave number, $k_{x,z}$ are the wave numbers along the $x,y$-axes,
$k_z$ is the far-field wave number in the direction orthogonal to the plane of the structure,
$\beta$ is the Bloch wave number, and, finally, $\ n=0,1, \ldots$ corresponds to the diffraction
order. Here, we consider $TM$-modes with $k_y=0$, i.e. propagating only perpendicular to the ridges, however,
a generalization to bi-directional BICs propagation in both $x,y$-directions is possible \cite{Bulgakov17a}.

For numerical simulations we use the rigorous Fourier modal approach in which the solution  the $y$-component of
the electric vector $E_y$ is written in the following form \cite{Pisarenco10,Pisarenco11}
\begin{equation}
E_y(x,z)=\sum_{n=-\infty}^{\infty}S_{n}(z)e^{-ik_{x,n}x}.
\end{equation}
The Fourier components $S_{n}(z)$ are matched on all interfaces
to cast Maxwell's equations into a set of linear equations truncated in the diffraction order.
Since the BICs are source-free solutions, our numerical implementation is restricted
to finding the poles of the scattering matrix. All simulations are run for
\begin{equation}\label{range}
\beta{a} < k_0{a} < 2\pi -\beta{a},
\end{equation}
which according to Eq.(\ref{spectrum}) means that only one TM scattering channel
is open in the far-zone on both sides of the DG. We mention in passing that a similar
couple wave approach was used in \cite{Gao16} for finding BICs in photonic crystal slabs
with one dimensional periodicity.

\section{Results}
The results of our simulation for DGs with $w=0.5a$ are collected in Table \ref{Table}. Our analysis of the numerical results showed that
in regard to BIC holding capacity to major types of DGs can be distinguished.
In the case of {\it asymmetric gratings} the system possesses no symmetry involving
mirror operation with respect to $x0y$-plane. This is always the case for
the different thicknesses of the upside and underside ridges $h_1 \neq h_2$. For example,
the DG shown in Fig \ref{Fig1} (a) clearly falls within this category. On the contrary, if
$h_1 = h_2$, and $\delta=0$ the system is mirror symmetric around its middle plane.
Another type of {\it symmetric gratings} are those possessing a glide symmetry i.e. a composition
of a mirror reflection and a half period translation along the $x$-axis with $\delta=0.5a$. In what follows we
discuss the specific features of BICs in the both types of DGs.
\begin{table}[t]
\begin{center}
\caption{BICs in DH for $w=0.5a$.}\label{Table}
\begin{tabular}{lccccccccc}
BIC & $k_0a$ & $\beta a$& $\delta$ & $L/a$&  $h_1/a$ & $h_2/a$ & $\epsilon_1$ & $\epsilon_2$\\
\hline
1     &  4.829 & 0        & 0     & 0.1747   & 1   & 0.5     &  1.5         & 3        \\
2     &  4.101 & 1.472    & 0     & 0.1747   & 1   & 0.5     &  1.5         & 3        \\
3     &  4.168 & 0        & N/A    & 0.8838  & 1     & 0       &  15          & 15        \\
4     &  4.221 & 1.311    & N/A    & 0.8838  & 1     & 0       &  15          & 15        \\
5     &  4.168 & 0        & 0.5    & 0.5248  & 1     & 0.5     &  1.5         & 3         \\
6     &  3.916 & 1.247    & 0.5    & 0.5248  & 1     & 0.5     &  1.5         & 3         \\
7     &  3.644 & 2.001    & 0      & 1       & 0.5   & 0.5     &  1.5         & 3         \\
8     &  5.094 & 0.690    & 0.5    & 1       & 0.5   & 0.5     &  1.5         & 3         \\
\end{tabular}\label{Table}
\end{center}
\end{table}

\subsection{Asymmetric gratings}
\begin{figure}[t]
\begin{center}
\includegraphics[width=.45\textwidth, height=0.73\textwidth,
trim={2.1cm 0.85cm 3.1cm 1.3cm},clip]{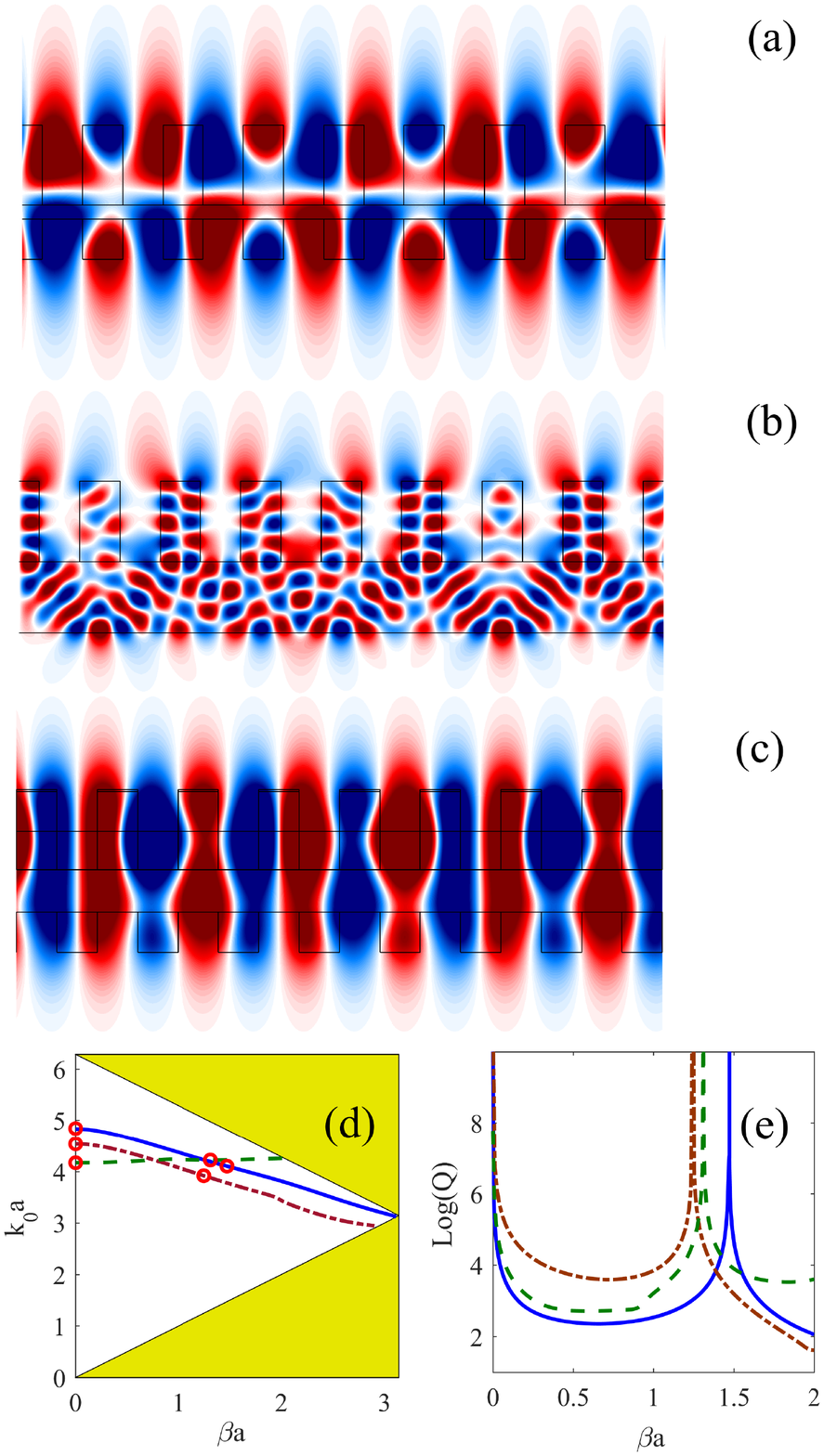}
\caption{BICs in asymmetric DGs. (a,b,c) Mode profiles in form of the real part of the electric vector
component $E_y$ for BICs 2,4,6 from Table \ref{Table}, respectively. (d) Dispersion of the real part of the leaky-mode
frequency: BIC 2 - solid blue; BIC 4 - dash green; BIC 6 dash-dot brown. The white domain is given by Eq.(\ref{range}). Red circles show the positions
of the BICs. (f) Q-factor dispersion for the same leaky
zones as in (d).  }\label{Fig2}
\end{center}
\end{figure}
The BICs in asymmetric DGs are BICs 1-6 from Table \ref{Table}. Among those BICs 2,4,6 are Bloch waves
with non-zero wave vector. The mode profiles of BICs 2,4,6 are shown in Fig \ref{Fig1}(a-c). At the
same time BICs 1,3,5 are symmetry protected standing wave BICs previously know in literature (not shown
here for brevity). Generally, the spectrum of a DG above the line of light
is characterized by leaky-modes \cite{Monticone15, Bulgakov16}, complex eigenfrequency dispersion branches
each of which can host a BIC in an exceptional point were the eigenfrequency is real and the Q-factor diverges
to infinity. In Fig. \ref{Fig2}(d) we shown the real part of the leaky-mode eigenfrequencies for three sets
of the DG parameters corresponding to BICs 2,4,6. One can see that besides a Bloch BIC
every dispersion branch also hosts a symmetry protected standing wave BIC in the $\Gamma$-point. These standing
wave BICs are BICs 1,3,5 from Table \ref{Table}. The dispersion of the Q-factors is shown in Fig. \ref{Fig2} (d)
to demonstrate its divergence in the points of BICs. It worth noting that the dispersion is symmetry with respect
$\beta \rightarrow -\beta$. Thus, each dispersion branch hosts to Bloch BICs propagating in the opposite directions.

One important feature of the asymmetric DG observed in numerical simulations is that finding a BIC always
requires tuning one of the systems's parameters. For instance, in our case the substrate thickness
$L$ always had to adjusted to find leaky zones with diverging Q-factor, as seen from Table \ref{Table}. Given that
the other parameters remain the same and the thickness is even slightly detuned from the values in Table \ref{Table}
the BICs disappear from the system. That feature will be explained later in the text.

\subsection{Symmetric gratings}
\begin{figure}[t]
\begin{center}
\includegraphics[width=.49\textwidth, height=.76\textwidth,
trim={5cm 3.0cm 5cm 8.6cm},clip]{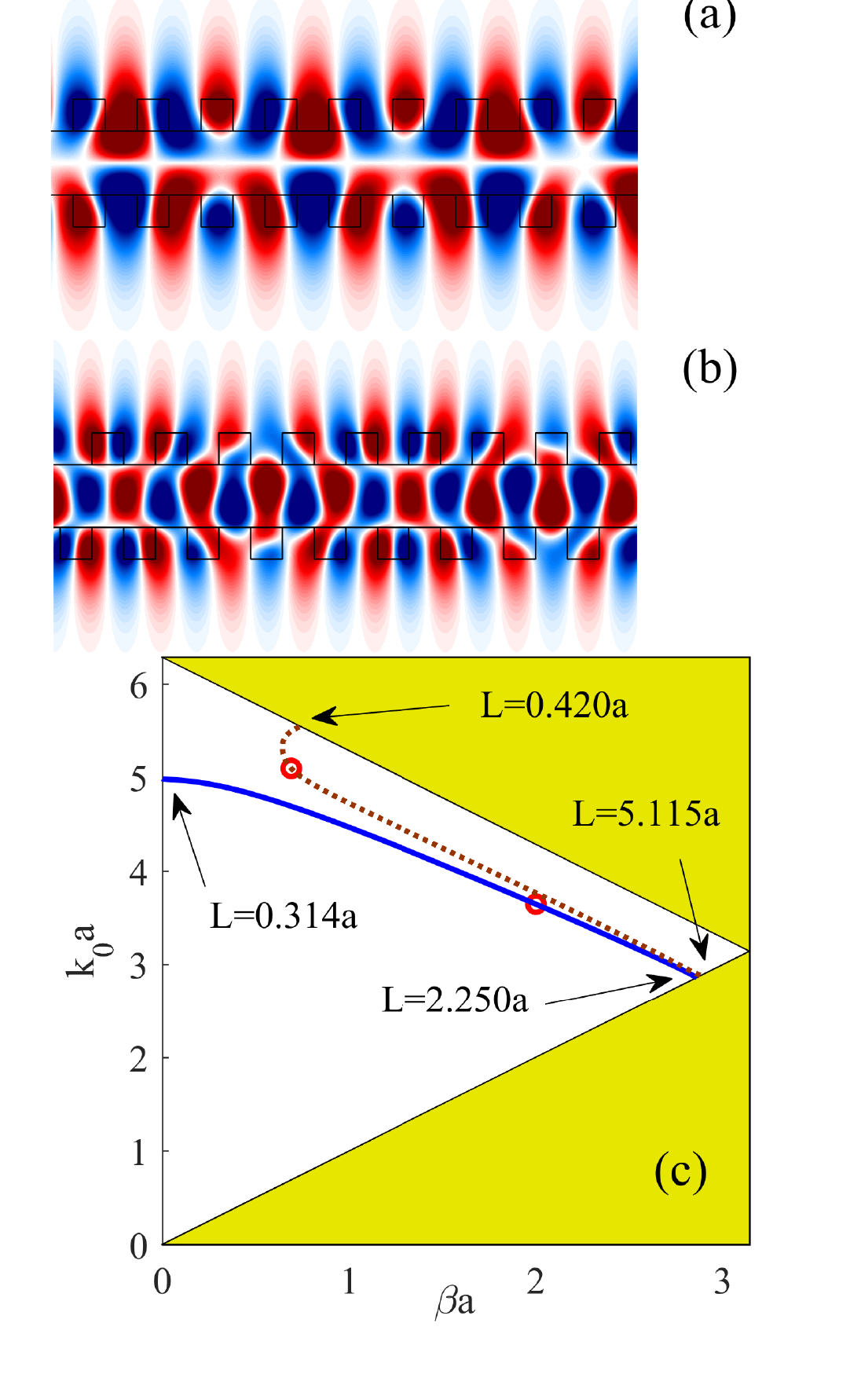}
\caption{BICs in symmetric DGs.(a,b) Mode profiles in form of the real part of the electric vector
component $E_y$ for BICs 7,8 from Table \ref{Table}, respectively. (c) Position of BIC 7 - solid blue, and BIC 8
dotted brown under variation of the substrate thickness. The white domain is given by Eq.(\ref{range}).}\label{Fig3}
\end{center}
\end{figure}

BICs 7,8 are supported by symmetric DGs. In the case of BIC 7 the DG has a mirror symmetry, while in the case
of BIC 7 the DG is glide symmetric. The mode profiles of BICs 7,8 are shown in Fig. \ref{Fig3}(a-b).
In contrast to asymmetric DGs, now finding a BIC does not require a fine tuning of the system's parameters.
This finding complies with the results presented in \cite{Ndangali2010} for double arrays of
infinitely thin dielectric rods. Moreover, if a control parameter, such as $L$, is slightly perturbed, the BIC
persists only having slightly different frequency $k_0$ and wave vector $\beta$.

This feature can be explained in a view of the topological properties of BICs in planar
structures, where the BICs are known to be associated with polarization singularities (vortices)
of the leaky zone far-field
polarization directions  \cite{Zhen2014,Bulgakov17}. Since the polarization singularity is topologically stable
the variation of parameters only relocates the position of polarization vortex in momentum space. Once
the DG is symmetric, the leaky modes also possess identical far-field patterns in upper and lower half-spaces that
can only be shifted with respect to each other in case of the glide symmetry. Therefore, under variation
of parameters the polarization vortex migrates in the same point in both upper and lower half-spaces ensuring
the stability of the BIC. This, however, is not the case for asymmetric DG when the BIC field pattern is also
asymmetric. Once the control parameter is perturbed the polarization vortices are relocated in momentum-space
in the both upside and underside far-field polarization patterns. Due to the absence of symmetry the positions
of the vortices do not have to coincide. Thus, Bloch BICs in asymmetric DGs are purely accidental in nature which
explains their fragility to variation of parameters.

Finally, let us illustrate the above arguments with numerical data. In Fig. \ref{Fig3}(c) we show
the frequencies of the families of BICs generated by BIC 7, and 8 under variation of the
substrate thickness $L$. It is seen from Fig. \ref{Fig3}(c) that with increase of $L$ both BICs shift to
the line of light $k_0=\beta$ until they eventually cross it to become ordinary guided modes below the line
of light protected by total internal reflection. With the decrease of $L$ the scenarios are, however, different.
The family generated by BIC 8 terminates at the line $k_0a=2\pi-\beta{a}$ which is the boundary of the second radiation
continuum. Once that boundary is crossed the BIC is destroyed by leakage to the second radiation channel.
In contrast to the above case, the family BIC 7 migrates to the $\Gamma$-point where all three BICs
hosted by the leaky zone coalesce. In this a pure Bloch BIC is destroyed giving rise to BICs propagating
along the ridges (see \cite{Bulgakov17} for more detail).

\section{Conclusion}
We considered Bloch bound states in the continuum in dielectric gratings.
Based on the Fourier modal approach we recovered the
leaky zones above the line of light to identify the
geometries of the gratings supporting Bloch bound states propagating in the direction perpendicular to
the ridges.
It shown that the capacity of dielectric gratings to host such bound states depends
on the presence/absence of symmetry with respect to the central plane of the grating.
It is demonstrated that if a two-side grating possesses either mirror or
glide symmetry
the Bloch bound states are stable to variation of parameters as far as the above symmetries are preserved.
That makes the bound states robust against possible fabrication inaccuracies at the same time allowing for a
certain
freedom in choosing the geometric parameters of the gratings. We speculate that our finding might be useful
in design of multifunction optical elements which steer the flow of light
harvested from the ambient medium.

\section*{Funding Information}

Ministry of Education and Science of Russian Federation
(state contract N3.1845.2017/4.6)


%

\bibliography{BSC_light_trapping}



\end{document}